\documentclass[12pt,a4paper,oneside,final,tilepage,onecolumn,thmsal]{article}
\usepackage[dvips]{graphics}
\usepackage{epsfig}
\usepackage{amsmath}

\newcommand{\QED}{\hspace*{\fill}\rule{2.5mm}{2.5mm}}

\usepackage{graphics}
\begin{document}
\def\beq{\begin{equation}}
\def\eeq{\end{equation}}
\def\bea{\begin{eqnarray}}
\def\eea{\end{eqnarray}}
\def\ve{\vert}
\def\vel{\left|}
\def\ver{\right|}
\def\nnb{\nonumber}
\def\ga{\left(}
\def\dr{\right)}
\def\aga{\left\{}
\def\adr{\right\}}
\def\rar{\rightarrow}
\def\nnb{\nonumber}
\def\la{\langle}
\def\ra{\rangle}
\def\ba{\begin{array}}
\def\ea{\end{array}}
\def\tep{$B \rar K \ell^+ \ell^-$}
\def\tepm{$B \rar K \mu^+ \mu^-$}
\def\tept{$B \rar K \tau^+ \tau^-$}
\def\ds{\displaystyle}
\title{{\small {\bf  Large-order shifted $1/N$ expansions through the asymptotic iteration method}}}
\author{\vspace{1cm}\\
{\small T. Barakat} \thanks {electronic address:
tbarakat@ksu.edu.sa}\\ {\small Physics Department, King saud University},\\
{\small Riyadh 11451, Saudi Arabia} }
\date{}
\begin{titlepage}
\maketitle
\thispagestyle{empty}
\begin{abstract}
\baselineskip .8 cm The perturbation technique within the
framework of the asymptotic iteration method is used to obtain
large-order shifted $1/N$ expansions, where $N$ is the number of
spatial dimensions. This method is contrary to the usual
Rayleigh-Schr\"{o}dinger perturbation theory, no matrix elements
need to be calculated. The method is applied to the
Schr\"{o}dinger equation and the non-polynomial potential
$V(r)=r^2+\frac{b r^2}{(1+cr^2)}$ in three dimensions is
discussed as an illustrative example.

\end{abstract}
\vspace{1cm} PACS number(s): 03.65 Ge
\end{titlepage}
\section{{\small Introduction}}
\baselineskip .8cm \hspace{0.6cm}

The shifted $1/N$ expansion technique (SLNT) proposed by Sukhatme
and Imbo [1] is an extremely powerful method of solving
Schr\"{o}dinger equation, and has been used extensively to
determine the eigenenergies for some important potentials [2-6].
The SLNT is an expansion in powers of $\Lambda^{-1/2}$,
$\Lambda=N+2\ell-a$, where $N$, $\ell$ and $a$ stand for the
number of dimensions, the angular momentum quantum number, and $a$
is a properly chosen shift parameter, respectively. The shift
parameter $a$ is usually chosen so as to improve the convergence
of the energy perturbation series, and to obtain the correct
answer for the harmonic oscillator and the hydrogen atom to all
orders.

After expanding the potential-energy function and the centrifugal
term in Taylor series about an appropriate point $r_{0}$ one is
left with the Hamiltonian operator for a harmonic oscillator plus
a polynomial perturbation. Then one applies perturbation theory
and obtains the perturbation corrections for the eigenfunctions
and eigenenergies.

The Rayleigh-Schr\"{o}dinger and the logarithmic perturbation
schemes (referred to as RSPT and LPT respectively) have been used
for the calculation of these corrections.

As well known the applications of the above two schemes were
restricted by serious difficulties. They require considerable
computational time and effort, and they involve, in general,
quite elaborate algebraic manipulations, so it was extremely
laborious to advance beyond the first four perturbation expansion
terms in the eigenenergy series [2].

On the other hand, considerable progress was made in the last few
years to obtain large-order shifted $1/N$ expansions [7-10]. For
example, Maluendes et al. [7] reported an approach in which the
coefficients of the SLNT of arbitrarily high orders could be
generated by means of the hypervirial (HV) and Hellmann-Feynman
theorems (HF), and thereby providing an excellent check for the
convergence of the method.

However, the previous authors in their work did not give explicit
expressions of their algorithm, each order getting progressively
much more complicated than the previous one, and the derivations
were tediously long. Thus, the need arises here to have a
relatively simple, fast and effective method that provides
large-order shifted $1/N$ expansions.

The so-called perturbation technique within the framework of the
asymptotic iteration method (AIM) [11] have emerged in recent
years to be a very useful and powerful technique of attack. This
method reproduced excellent results for many important potentials
in relativistic and non-relativistic quantum mechanics. Through
AIM one can actually obtain all the perturbation corrections to
both energy level shifts and wavefunctions for all states [12,
13]. These quantities can be calculated to any given accuracy,
since the generation of successive corrections in the present
perturbative framework, only requires the solution of simple
algebraic equation.

The method is also applicable in the same form to both the ground
state and excited bound states without involving tedious
calculations which appeared in the available perturbation
theories.

Encouraged by its satisfactory performance, we feel tempted to
extend AIM, and to see this time how the AIM can be used to
obtain large-order shifted $1/N$ expansions for the
three-dimensional Schr\"{o}dinger equation with any arbitrary
spherically symmetric potential $V(r)$ directly without either,
using the base eigenfunctions of the unperturbed problem, or
needing to calculate matrix elements.

As an illustration, the present technique is applied to the
non-polynomial potential $V(r)=r^2+\frac{b r^2}{(1+cr^2)}$. This
potential appears in several areas of physics. In field theory it
provides a simple zero-dimensional model possessing a
non-polynomial Lagrangian [14]. In laser physics it arises out of
the Fokker-Planck equation for a single-mode laser [15].

For this potential only a class of exact analytical solutions for
certain parameter dependence $b=b(c)$ were obtained [16, 17].
Hence, it has been a subject of several investigations and many
authors have studied the one-, two-and three-dimensional cases
[18, 19]. Roy, Roychoudhury and Roy have shown the supersymetric
character of this potential and given new solutions using the
standard $1/N$ expansion method [20].

With this in mind, this paper is organized as follows. In Sec. 2,
the formulation of SLNT through AIM is outlined to find the
eigenenergies for any arbitrary spherically symmetric potential.
The analytical expressions for AIM are cast in such a way that
allows the reader to use them without proceeding into their
derivation. In Sec. 3, we explained how to obtain numerically the
eigenenergies for the non-polynomial potential $V(r)=r^2+\frac{b
r^2}{(1+cr^2)}$, and therein we will compare the accuracy of our
results with those obtained by Roy et al. [20]. Finally, the paper
ends with a brief summary and concluding remarks on the method
and our findings.

\section{{\small Formalism of the asymptotic iteration method for SLNT}}
\hspace{0.1cm} The radial part of the time-independent
Schr\"odinger equation for central-field model in terms of the
expansion parameter $\Lambda$ in $N$- dimensional space with
$(\hbar=2m=1)$ is
\begin{eqnarray}
\left\{-\frac{d^{2}}{dr^{2}}+ \frac{\Lambda^2}{4
r^{2}}(1+\frac{2A}{\Lambda}+\frac{4B}{\Lambda^2})\right\}\chi_{n\ell}(r)=E_{n\ell}\chi_{n\ell}(r),
\end{eqnarray}
with

$\Lambda=N+2\ell-a$, ~~~~$A=1-N+a$, ~~~~$B=(N-a)(N-a-2)/4$.

SLNT begins with shifting the origin of the coordinate through the definition

\begin{eqnarray}
x=\Lambda^{1/2}(\frac{r}{r_{0}}-1),
\end{eqnarray}
where $r_{0}$ is chosen to minimize the effective potential
$V_{eff}(r)=\Lambda^{2}(\frac{1}{4r^2}+\frac{V(r)}{\Lambda^{2}})$,
so that
\begin{eqnarray}
\Lambda^2=2r_{0}^{3}V'(r_{0}).
\end{eqnarray}

Expansions about $r=r_{0}$, yield

\begin{eqnarray}
\left\{-\frac{d^{2}}{dx^{2}}+\sum_{i=0}^{\infty}(\alpha_{i}g^{i}x^{i+2}+
\beta_{i}g^{i}x^{i}+\xi_{i}g^{i+2}x^{i})\right\} \chi_{n\ell}(x)=\epsilon_{n\ell}\chi_{n\ell}(x)
\end{eqnarray}
where
\begin{eqnarray}
\alpha_{i}=(-1)^{i}\frac{i+3}{4}+\frac{r^{i+1}_{0}}{2(i+2)!}\frac{d^{i+2}V(r_{0})}{V^{'}(r_{0})dr_{0}^{i+2}},
{~}\beta_{i}=(-1)^{i}\frac{i+1}{2}A,{~}\xi_{i}=(-1)^{i}(i+1)B,
\end{eqnarray}
and
\begin{eqnarray}
\epsilon_{n\ell}=r_{0}^2g^2(E_{n\ell}-V_{eff}(r_{0})),~{\rm
and}~~g=1/\Lambda^{1/2}.
\end{eqnarray}

The systematic procedure of the AIM begins now by expanding the
eigenenergy term in powers of $g$, and rewriting equation (4) in
the following form
\begin{eqnarray}
&&\left[-\frac{d^{2}}{dx^{2}}+\alpha_{0}x^{2}+\beta_{0}+g(\alpha_{1}x^{3}+
\beta_{1}x)+g^{2}(\alpha_{2}x^{4}+\beta_{2}x^2+\xi_{0})
\right. \nonumber \\
&&\nonumber\\
&&\left.
+g^3(\alpha_{3}x^{5}+\beta_{3}x^{3}+\xi_{1}x)+g^4(\alpha_{4}x^{6}+\beta_{4}x^{4}+\xi_{2}x^2)+\cdots
\right]\chi_{n\ell}(r)
\nonumber \\
&&\nonumber \\
&&=\left[\epsilon^{(0)}_{n\ell}+g\epsilon^{(1)}_{n\ell}+g^{2}\epsilon^{(2)}_{n\ell}+g^3\epsilon^{(3)}_{n\ell}+
g^4\epsilon^{(4)}_{n\ell}+\cdots \right]\chi_{n\ell}(x).
\end{eqnarray}

If we further insert the ansatz
\begin{eqnarray}
\chi_{n\ell}(x)=e^{-\gamma x^{2}/2}f_{n\ell}(x)
\end{eqnarray}
into equation (7), carrying out the mathematics, in this case, the
function $f_{n\ell}(x)$ will satisfy a new second-order
homogeneous linear differential equation of the form
\begin{eqnarray}
f_{n\ell}^{''}(x)=\lambda_{0}(x,g)f_{n\ell}^{'}(x)+s_{0}(x,g)f_{n\ell}(x),
\end{eqnarray}
where $\lambda_{0}(x,g)=2\gamma x$, and
\begin{eqnarray}
s_{0}(x,g)&=&(\alpha_{0}-\gamma^{2})x^2+\beta_{0}+\gamma-
\epsilon_{n\ell}+g(\alpha_{1}x^{3}+\beta_{1}x)+g^2(\alpha_{2}x^{4}+\beta_{2}x^2+\xi_{0})\nonumber\\
&+&g^3(\alpha_{3}x^{5}+\beta_{3}x^{3}+\xi_{1}x)+g^4(\alpha_{4}x^{6}+\beta_{4}x^{4}+\xi_{2}x^2)+\cdots,
\end{eqnarray}
\begin{eqnarray}
\epsilon_{n\ell}=\epsilon^{(0)}_{n\ell}+g
\epsilon^{(1)}_{n\ell}+g^2 \epsilon^{(2)}_{n\ell}+g^3
\epsilon^{(3)}_{n\ell}+g^4 \epsilon^{(4)}_{n\ell}+.....
\end{eqnarray}

Here, it should be pointed out that, the choice of $g$ in equation
(9) will be motivated, that is when we switch off $g$, equation
(9) will be reduced to an exactly solvable eigenvalue problem
within the framework of AIM [21-23].

To apply the perturbation expansion technique within the
framework of AIM we rely on the symmetric structure of the right
hand side of equation (9). Thus, we differentiate equation (9)
$(k +2)$ times with respect to $x$, $k = 1, 2, . . .$. Then we
take the ratio of the $(k + 2)^{th}$ {\rm and} $(k + 1)^{th}$
derivatives, and for sufficiently large $k$ we introduce,
respectively, the "asymptotic" aspect and the termination
condition of the method, which, in turn will lead to
\begin{eqnarray}                                    %14
\varrho(x,g)\equiv\frac{s_k(x,g)}{\lambda_k(x,g)}=\frac{s_{k-1}(x,g)}{\lambda_{k-1}(x,g)},
\end{eqnarray}
\begin{eqnarray}
\delta_{k}(x,g)\equiv
s_k(x,g)\lambda_{k+1}(x,g)-s_{k+1}(x,g)\lambda_{k}(x,g)=0.
\end{eqnarray}
We now proceed to obtain the eigenenergies of equation (9)
systematically in terms of the expansion parameter $g$. If we
expand $\delta_{k}(x,g)$ around $g =0$, we get the following
series
\begin{eqnarray}
\delta_{k}(x,g)=\delta_{k}(x,0)+\frac{g}{1!}\frac{\partial\delta_{k}(x,g)}{\partial
g}|_{g=0}+ \frac{g^2}{2!}\frac{\partial^2\delta_{k}(x,g)}{\partial
g^2}|_{g=0}+
\frac{g^3}{3!}\frac{\partial^3\delta_{k}(x,g)}{\partial
g^3}|_{g=0}+....
\end{eqnarray}
According to the procedure of AIM [21-23], $\delta_{k}(x,g)$ must
be zero; if this to be true for every $g$ value, then every term
of the series must be zero. That is to say
\begin{eqnarray}
\delta^{(j)}_{k}(x,g)=\frac{g^j}{j!}\frac{\partial^{j}\delta_{k}(x,g)}{\partial
g^{j}}|_{g=0}=0, ~~j=0,1,2....
\end{eqnarray}

A quantitative estimate for $\epsilon_{n\ell}$ expansion terms can
be obtained by comparing the terms with the same order of $g$ in
equations (9) and (14). Therefore, it is clear that the roots of
$\delta^{(0)}_{k}(x, 0)=0$ give us the zero'th contribution energy
terms $\epsilon^{(0)}_{n\ell}$. Likewise, the roots of
$\delta^{(1)}_{k}(x,g)|_{g=0}=0$ give us the first correction
terms $\epsilon^{(1)}_{n\ell}$, and so on. Therefore, the general
solution for the eigenenergies $E_{n\ell}$ in conjunction with
equations (6) and (11) is
\begin{eqnarray}
E_{n\ell}=\frac{\Lambda^2}{r^2_{0}}\left(\frac{1}{4}+\frac{r^{2}_{0}V(r_{0})}{\Lambda^2}\right)
+\frac{1}{r^{2}_{0}g^2}\sum_{i=0}^{\infty}g^{i}\epsilon^{(i)}_{n\ell}.
\end{eqnarray}

\section{{\small Numerical results for the eigenenergies of
the potential $V(r)=r^2+\frac{b r^2}{(1+cr^2)}$ }}

Within the framework of the asymptotic iteration method mentioned
in the above section, the eigenenergies $E_{n\ell}$ of the
non-polynomial potential $V(r)=r^2+\frac{b r^2}{(1+cr^2)}$ are
calculated by means of equation (16).

To obtain the zero'th contribution energy terms
$\epsilon^{(0)}_{n\ell}$, one should simply switch off $g$ in
equation (9), that will lead to an exactly solvable eigenvalue
problem within the framework of AIM,
\begin{eqnarray}
f_{n\ell}^{''}(x)=2\gamma x
f_{n\ell}^{'}(x)+\left((\alpha_{0}-\gamma^{2})x^2+\beta_{0}+\gamma-
\epsilon^{(0)}_{n\ell}\right)f_{n\ell}(x).
\end{eqnarray}
For each iteration, the expression $\delta^{(0)}_{k}(x,0)=0$ in
equation (13) depends on two variables namely $\gamma$ and $x$.
Since the problem is exactly solvable, the calculated
eigenenergies $\epsilon^{(0)}_{n\ell}$ by means of this condition
are independent of the choice of $x$ once we set
$\gamma=\sqrt{\alpha_{0}}$; then the roots of
$\delta^{(0)}_{k}(x,0)=0$ are
\begin{eqnarray}
\epsilon^{(0)}_{n\ell}=\beta_{0}+(2n+1)(\alpha_{0})^{1/2},~~~~n=0,1,2,....
\end{eqnarray}

As we noted before, the leading contribution term of the total
energy is of order $\Lambda^2$ given in equation (16). The next
contribution is of order $\Lambda$ and is given by
$\beta_{0}+(2n+1)(\alpha_{0})^{1/2}$. It is customary to choose
the shift parameter $a$ so as to make this contribution vanish.
This choice is physically motivated by requiring agreement
between the $1/\Lambda$ expansions and the exact analytic results
for the harmonic-oscillator and Coulomb potentials to all orders
[2]. However, there are some difficult cases in which this simple
choice is insufficient and it is really necessary to select an
order-dependent value of $a$ according to minimal sensitivity or
other appropriate criterion [7]. Nevertheless, at this point, it
is enough for us to compare the AIM results with the SLNT results.
Therefore, we choose
\begin{eqnarray}
a=2-2(2n+1)(\alpha_{0})^{1/2},
\end{eqnarray}
$\Lambda=N+2\ell-a$, and $\Lambda^2=2r_{0}^{3}V'(r_{0})$.
Collecting these terms and carrying out the mathematics, one can
get
\begin{eqnarray}
3+2\ell-2+2(2n+1)(\alpha_{0})^{1/2}=(2r_{0}^{3}V'(r_{0}))^{1/2},
\end{eqnarray}
which is an implicit equation for $r_{0}$. Once $r_{0}$ is
determined, the leading term $\Lambda^{2}V_{eff}(r_{0})$ can be
calculated numerically. On the other hand, to obtain the higher
order perturbative expansion terms, first one should go to
equation (9) switch on $g$ and then replace $\epsilon_{n\ell}$
with $\epsilon^{(0)}_{n\ell}+g \epsilon^{(1)}_{n\ell}$, and
terminate the iterations by imposing the condition
$\delta^{(1)}_{k}(x,g)$= 0 as an approximation to equation (9).
The first root of the resulting equation gives
$\epsilon^{(1)}_{n\ell}$. Similarly, and very easily one can
obtain the other perturbative expansion terms.

Throughout the present calculations, it is observed that the
perturbation corrections of odd orders
$\epsilon^{(2i+1)}_{n\ell}$ are vanish for all $i$.

In table 1 an explicit list of calculations up to six'th order
with different values of $\ell$, $b$, and $c$ are given, so that
the reader may, if so inclined, reproduce our results. In table 2
the results of AIM, together with the standard shifted expansion
method $E_{n\ell}(1/N)$ and the exact super-symmetric results are
displayed for comparison purposes.

In tables we have only considered the eigenenergies for the
ground state $n=0$. This was in order to make a clear comparison
between the results of this method and the results of [20].
Examination of tables shows that the accuracy of the AIM is
better than the accuracy of shifted 1/N expansion method, and the
predicted eigenenergies $E_{n\ell}$(AIM) are all in excellent
agreement with the results of the super-symmetric method [20].

We have also shown that, it is very easy task to implement the
perturbation technique within the framework of the AIM without
having to be worry about the ranges of the couplings in the
potential.

This method is also applicable in the same form to both the ground
state, and excited bound states without involving tedious
calculations which appeared in the available perturbation
theories.

As a concluding remark, the present method enable one if it is
necessary, to keep $a$ as a free parameter up to the end of the
calculations. With this choice the results can be drastically
improved by raising up the perturbative order in the expansion to
any order, and then one can determine the value of $a$ according
to the minimal sensitivity method or any other appropriate
criterion.

\clearpage
\begin{tiny}
\begin{table}
\begin{center}
\caption{The calculated values of the coefficients in the energy
expansion $\epsilon_{n\ell}$ for the non-polynomial potential by
means of this work with different values of $\ell$, $b$, and $c$.}
\vspace{1cm}
\begin{tabular}{cccccccccc}
\hline \hline $\ell$ &c&b &$\epsilon^{(0)}_{0\ell}$
&$\epsilon^{(1)}_{0\ell}$&$\epsilon^{(2)}_{0\ell}$&$\epsilon^{(3)}_{0\ell}$
&$\epsilon^{(4)}_{0\ell}$ &$\epsilon^{(5)}_{0\ell}$ &$\epsilon^{(6)}_{0\ell}$ \\
\hline \hline
0  &0.1  &-0.46    &0  &0      &-0.021504914    &0    &0.01472806       &0       &-0.010867      \\
1  &0.1  &-0.5     &0  &0      &-0.030894587    &0    &0.02042523       &0       &-0.0034827     \\
2  &0.1  &-0.54    &0  &0      &-0.034627947    &0    &0.01793987       &0       & 0.018511       \\
0  &0.01 &-0.0406  &0  &0      &-0.000018842    &0    &1.707706x$10^-6$ &0       &-1.7717x$10^-6$ \\
1  &0.01 &-0.041   &0  &0      &-0.000049912    &0    &7.312060x$10^-6$ &0       &-7.6296x$10^-6$ \\
-1 &0.1  &-0.42    &0  &0      &-0.005539564    &0    &0.002341067      &0       &-0.0018839       \\
\hline \hline
\end{tabular}
\end{center}
\end{table}
\end{tiny}
\clearpage
\begin{tiny}
\begin{table}
\begin{center}
\caption{Comparison between selected eigenenergies calculated
from the standard shifted (1/N) expansion method $E_{n\ell}(1/N)$
[20], the exact super-symmetric values $E_{n\ell}(SUSY)$ [20],
and the eigenenergies $E_{n\ell}(AIM)$ computed by means of
equation (16) up to four'th and six'th orders.} \vspace{1cm}
\begin{tabular}{ccccccccc}
\hline \hline $\ell$ &c&b& $E_{0\ell}(1/N)$& $E_{0\ell}(SUSY)$&$E^{4th}_{0\ell}(AIM)$&$E^{6th}_{0\ell}(AIM)$\\
\hline \hline
0  &0.1   &-0.46   &2.400520  &2.4   &2.40051591814138       &2.3999024133815471            \\
1  &0.1   &-0.5    &4.000116  &4.0   &4.00012195111641       &4.0000777082899992              \\
2  &0.1   &-0.54   &5.599965  &5.6   &5.59998537049538       &5.6000735787721094              \\
0  &0.01  &-0.0406 &2.939999  &2.94  &2.94000001431155       &2.9399998857627638            \\
1  &0.01  &-0.041  &4.899974  &4.9   &4.90000002136554       &4.8999999017683162             \\
-1 &0.1   &-0.041  &0.801177  &0.8   &0.80117658318143       &0.7984848338375296          \\
\hline \hline
\end{tabular}
\end{center}
\end{table}
\end{tiny}

\clearpage
\begin{thebibliography}{99}
\bibitem{R1}Sukhatme U, and Imbo T 1983 Phys. Rev. D{\bf 28}
418
\bibitem{R2}Imbo T, Pagnamenta A, and Sukhatme U 1984 Phys. Rev. D{\bf 29}
1669
\bibitem{R3}Imbo T, Pagnamenta A, and Sukhatme U 1984 Phys. Lett. A{\bf 105}
183
\bibitem{R4}Imbo T, and Sukhatme U 1985 Phys. Rev. D {\bf 31}
2655
\bibitem{R5}Dutta R, Mukherji U, and Varshni Y P 1986 Phys. Rev. A{\bf 34}
777
\bibitem{R6}Roy B  1986 Phys. Rev. A{\bf 34} 5108

\bibitem{R7}Maluendes S A, Fern$\acute{a}$ndez F M, Mes$\acute{o}$ A M, and Castro E A 1986 Phys. Rev. D{\bf 34} 1835
\bibitem{R8}Chaterjee A 1990 Phys. Rep.{\bf 186} 249
\bibitem{R9}Fern$\acute{a}$ndez F M  2002 J. Phys. A{\bf 35} 10663
\bibitem{R10}Mustafa O  2002 J. Phys. A{\bf 35} 10671
\bibitem{R11}Ciftci H, Hall R. L and Saad N 2005 Phys. Lett. A{\bf340}
388
\bibitem{R12}Barakat T 2006 J. Phys. A{\bf 39} 823
\bibitem{R13}Barakat T 2006 Int. J. Mod. Phys. A{\bf 21} 4127
\bibitem{R14}Biswas S N, Datta K, Saxena R P, Srivastava P K, and
Varma V S 1973 J. Math. Phys. {\bf 14} 1190
\bibitem{R15}Risken H, and Vollmer H D 1967 Z. Phys. {\bf 201} 323
\bibitem{R16}Lai C S, and Lin H E 1982 J. Phys. A{\bf 15} 1495
\bibitem{R17}Flessas G P 1981 Phys. Lett. A{\bf 83} 121
\bibitem{R18}Pons R, and Marcilhacy 1991 Phys. Lett. A{\bf 152}
235
\bibitem{R19}Bose S K, and Varma N 1989 Phys. Lett. A{\bf 141} 141
\bibitem{R20}Roy B, Roychoudhury, and Roy P 1988 J. Phys. A{\bf 21}
1579
\bibitem{R21}Ciftci H, Hall R. L, and Saad N 2003 J. Phys.
A{\bf 36} 11807
\bibitem{R22}Fern$\acute{a}$ndez F M 2004 J. Phys.
A{\bf 37} 6173
\bibitem{R23}Barakat T, Abodayeh K, and Mukheimer A 2005 J. Phys.
A{\bf 38} 1299
\bibitem{R24}Barakat T 2005 Phys. Lett. A{\bf 344} 411
\end{thebibliography}
\end{document}